\documentclass[journal=jctcce,manuscript=article]{achemso}

\SectionNumbersOn

\usepackage[T1]{fontenc}
\usepackage{graphicx}
\usepackage{dcolumn}
\usepackage{bm}
\usepackage{mathrsfs}
\usepackage{amsmath}
\usepackage[linesnumbered,ruled,vlined]{algorithm2e}
\usepackage{float}
\usepackage{enumitem}
\usepackage{tabulary}
\usepackage{soul}
\usepackage{color}
\usepackage{xcolor}
\usepackage{multirow}
\usepackage{subfig}
\usepackage{caption}
\newsavebox{\bigleftbox}

\author{Yi Fan}
\affiliation{Hefei National Research Center for Physical Sciences at the Microscale, University of Science and Technology of China, Hefei, Anhui 230026, China}

\author{Jie Liu}
\affiliation{Hefei National Laboratory, University of Science and Technology of China, Hefei 230088, China}

\author{Zhenyu Li}
\email{zyli@ustc.edu.cn}
\affiliation{Hefei National Research Center for Physical Sciences at the Microscale, University of Science and Technology of China, Hefei, Anhui 230026, China}
\alsoaffiliation{Hefei National Laboratory, University of Science and Technology of China, Hefei 230088, China}

\author{Jinlong Yang}
\affiliation{Hefei National Research Center for Physical Sciences at the Microscale, University of Science and Technology of China, Hefei, Anhui 230026, China}
\alsoaffiliation{Hefei National Laboratory, University of Science and Technology of China, Hefei 230088, China}

\title{Quantum circuit matrix product state ansatz for large-scale simulations of molecules}

\begin{document}
	\maketitle
	
	 \begin{tocentry}
		\begin{center}
			\includegraphics[height=4.3cm]{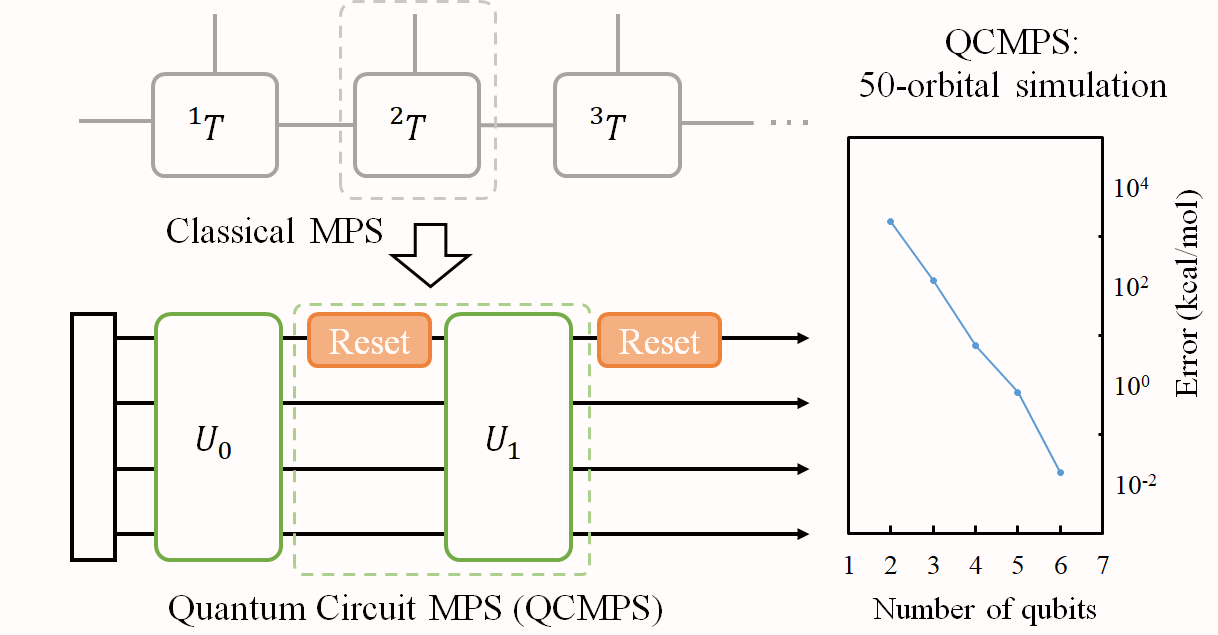}
		\end{center}
	\end{tocentry}

    \begin{abstract}
       As in the density matrix renormalization group (DMRG) method, approximating many-body wave function of electrons using a matrix product state (MPS) is a promising way to solve electronic structure problems. The expressibility of an MPS is determined by the size of the matrices or in other words the bond dimension, which unfortunately should be very large in many cases. In this study, we propose to calculate the ground state energies of molecular systems by variationally optimizing quantum circuit MPS (QCMPS) with a relatively small number of qubits. It is demonstrated that with carefully chosen circuit structure and orbital localization scheme, QCMPS can reach a similar accuracy as that achieved in DMRG with an exponentially large bond dimension. QCMPS simulation of a linear molecule with 50 orbitals can reach the chemical accuracy using only 6 qubits at a moderate circuit depth. These results suggest that QCMPS is a promising wave function ansatz in the variational quantum eigensolver algorithm for molecular systems.
    \end{abstract}

	\section{Introduction}
	
	Electronic structure theory is essential in studying properties of chemical systems. While classical methods such as configuration interaction (CI), M\o{}ller-Plesset perturbation theory and coupled cluster (CC) theory provide systematical ways to approach accurate ground states, the computational cost grows rapidly and quickly goes beyond the capability of current computers. Advances in quantum technology provide a promising pathway to solve the electronic structure problem, which is expected to be one of the fields to demonstrate practical quantum advantage in the near future\cite{Tilly_VQE_2021, McAEndAsp20, Cerezo2021, Magann_VQA_2021, CaoRomOls19, Fedorov_VQERev_2022}. By encoding the electronic wave function into the quantum state of qubits, the many-body Schr\"odinger equation can be solved using quantum algorithms such as quantum phase estimation (QPE) and variational quantum eigensolver (VQE).\cite{BraKit02, McAEndAsp20, CaoRomOls19, Pre18, GeoAshNor14, AspDutLov05, Wang08, PerMcCSha14, HemMaiRom18, NamChen20, SheZhaZha17, MalBabKiv16, Kandala2017, ColRamDah18, McCRomBab16, LanWhi10, RomBabMcC18, vqe-excited-vqd, Mcclean17qse, YungCas2014} Subject to limited qubit resources and noisy realizations of near-terms quantum devices\cite{Pre18, Boixo2018}, the quantum-classical hybrid VQE method is more preferable for noisy intermediate scale quantum (NISQ) devices. In recent years, variational quantum circuit ansatzes such as the unitary coupled-cluster (UCC)\cite{Kut82, BarKucNog89, TauBar06} and hardware-efficient ansatz (HEA)\cite{Kandala2017} have been successfully applied in quantum chemistry applications for small molecules on most leading quantum hardware platforms\cite{PerMcCSha14, HemMaiRom18, NamChen20, SheZhaZha17, MalBabKiv16, Kandala2017, ColRamDah18}. Circuit optimization techniques including operator reduction\cite{cao_SYMM_2022, Yordanov2021, Ryabinkin2021, Joonho2018, Bauman2019} and adaptive algorithms\cite{Grimsley2019, Tang2021, Ryabinkin_iQCC_2020, Ryabinkin_iQCC_2021, Liu_ADAPTx_2021, Liu_effHam_2022, Ratini_WAHTOR_2022, Fan_ESVQE_2021, Tsuchimochi_ADAPT_CIRC_2022, Dyke_ADAPT_POOL_2022, Zhang_ES_LIKE_2021, Eddins_ENT_FORG_2022, Burton_DISCO_VQE_2022} have also been proposed to further reduce circuit depth and lower optimization overhead.
	
    In most VQE ansatzes, an orbital-to-qubit mapping is implemented, where orbital occupations are mapped to $|0\rangle$ and $|1\rangle$ states of qubits. The number of qubits required to encode the eigen states of an electronic Hamiltonian is then determined by the number of basis functions, which can easily exceed the capacity of NISQ devices. The recently proposed quantum circuit tensor network (TN) ansatz provides a possible way to solve this problem\cite{Liu_FEW_QUBIT_2019, Haghshenas_QCTN_2022}. By encoding a classical TN state into a structured parametric quantum circuit with mid-circuit measurements, the quantum circuit TN ansatz is capable of generating tensors with an exponentially large bond dimension using a small number of qubits. Liu \textit{et al.} demonstrated that matrix product state (MPS) and projected entangled pair state (PEPS) represented by quantum circuits are capable of simulating ground states of the Heisenberg model with a high fidelity using much fewer qubits than the system size\cite{Liu_FEW_QUBIT_2019}. Haghshenas \textit{et al.} further investigated the variational expressibility of quantum circuit MPS 
    (QCMPS) against classical MPS and suggested that QCMPS can be more parameter-efficient than its classical counterpart\cite{Haghshenas_QCTN_2022}. Therefore, it is interesting to see whether it is possible to use the qubit-efficient QCMPS to simulate large molecules which are difficult to be handled by conventional VQE ansatzes due to the high qubit requirement and also by classical MPS methods, such as the density matrix renormalization group (DMRG) algorithm, \cite{White_DMRG_1992, White_DMRG_1993} due to the high bond-dimension requirement.

    Implementing QCMPS based VQE algorithms in molecular systems will face two challenges. A classical MPS is composed of dense matrices which are fully flexible, while in QCMPS each matrix is represented by a structured circuit block and the parameter space is thus constrained by the specific circuit structure. Such a sparse parameterization certainly has influences on its variational expressibility. Previous studies used nearest-neighbour entangled blocks, such as the SU(2) symmetry circuit structures\cite{Liu_FEW_QUBIT_2019} and the brick-wall structure\cite{Haghshenas_QCTN_2022}, which may be inefficient to capture electron correlations accurately. Therefore, the first challenge is to choose a suitable circuit structure to build QCMPS for molecular systems.
    On the other hand, although the number of qubits in QCMPS does not explicitly depend on the size of the system or more specifically the number of orbitals, simulating a larger system may still require a rapidly growing bond dimension thus more qubits to achieve certain accuracy due to the violation of the area law\cite{Eisert_AREA_LAW_2010} of electronic Hamiltonian. Therefore, the second challenge is to suppress the increase in bond dimension. Similar to orbital localization in DMRG, a proper way to change the basis for the electronic Hamiltonian is expected to be necessary in QCMPS.

	In this work, we developed a QCMPS ansatz for variational optimization of ground states of molecular systems. The accuracy of the QCMPS is investigated with different local circuit structures and molecular orbital basis. A linearly entangled structure is compared with the fully entangled blocks. The influence of orbital localization on the performance of QCMPS is analyzed. As a demonstration, QCMPS is successfully applied in simulating ground states of hydrogen chains with up to 50 orbitals, reaching chemical accuracy using a very small number of qubits. Results presented here open a new avenue for VQE simulation for large molecular systems.

    \section{Methods}
    
\subsection{Variational Quantum Eigensolver}
    \begin{figure}[tb]
    		\subfloat[]{\includegraphics[width=0.70\linewidth]{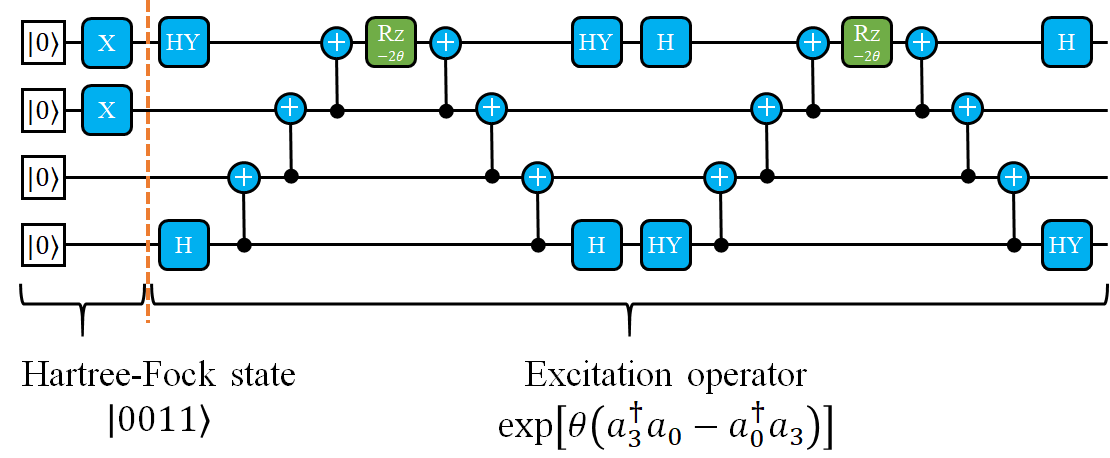}} \\
    		\subfloat[]{\includegraphics[width=0.70\linewidth]{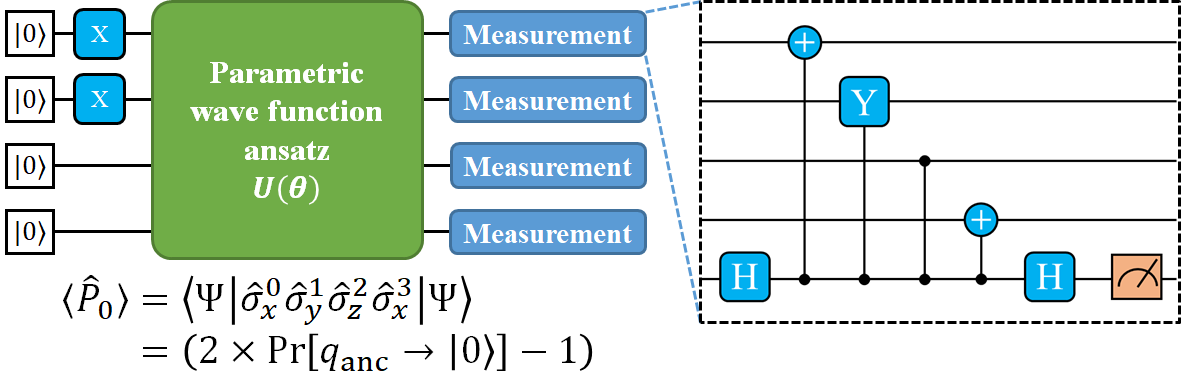}} \\
    		\caption{(a) A quantum circuit representing $\exp{[\theta(a^{\dagger}_{3} a_{0} - a^{\dagger}_{0} a_{3})]}|\Psi_{HF}\rangle$ under the Jordan-Wigner transformation. $\texttt{HY}$ is defined as $\texttt{HY}=\sqrt{2}/2\times (\texttt{Z}+\texttt{Y})$. (b) Hadamard test circuit to calculate the expectation value of $\hat{P}_{0}=\hat{\sigma}^{0}_{x}\hat{\sigma}^{1}_{y}\hat{\sigma}^{2}_{z}\hat{\sigma}^{3}_{x}$, where $q_{\text{anc}}$ stands for the bottom ancillary qubit.}
    		\label{fig::circ-cvqe}
    \end{figure}

    VQE computes expectation values of a given Hamiltonian on a quantum computer and performs variational optimization of wave function parameters using a classical algorithm. For a qubit Hamiltonian which is expressed as a linear combination of Pauli strings:
    \begin{equation}
        \hat{H} = \sum_{i}{c_{i} \hat{P}_{i}}
        \label{eq::qubit-ham}
    \end{equation}
    where $\hat{P}_{i}$ is product of Pauli operators $\{\hat{\sigma}_{x}, \hat{\sigma}_{y}, \hat{\sigma}_{z}, \hat{\sigma}_{I}\}$, the expectation value of each Pauli string can be obtained through quantum measurements, and the expectation value of $\hat{H}$ is thus expressed as a summation:
    \begin{equation}
        \begin{aligned}
            E(\theta) &= \langle \Psi(\theta) | \hat{H} | \Psi(\theta) \rangle = \langle \Psi(\theta) | \sum_{i}{c_{i} \hat{P}_{i}} | \Psi(\theta) \rangle \\
              &= \sum_{i}{c_{i} \langle \Psi(\theta) | \hat{P}_{i} | \Psi(\theta) \rangle}.
        \end{aligned}
        \label{eq::qubit-ham-detail}
    \end{equation}
    An important part of VQE is the encoding of target wave function $|\Psi(\theta)\rangle$ using a parametric quantum circuit. Conventional VQE ansatzes usually use an orbital-to-qubit mapping, which is a straightforward strategy to represent the eigen states of $\hat{H}$ using the quantum states of qubits. For example, under the Jordan-Wigner mapping\cite{JordanWigner_1928}, the orbital occupation is represented directly by the qubit states $|0\rangle$ and $|1\rangle$. As shown in Figure~\ref{fig::circ-cvqe}a, a two-electron Hartree Fock (HF) wave function $|0011\rangle$ can be simply obtained using a four-qubit quantum circuit by applying two \texttt{X} gates on the first two qubits, and the unitary excitation operator $\exp{[\theta(a^{\dagger}_{3} a_{0} - a^{\dagger}_{0} a_{3})]}$ corresponds to a W-shape circuit which is parameterized by the \texttt{Rz} gates. Using such a protocol, it is convenient to construct the parametric quantum circuit which encodes the wave function from orbital-based physically motivated ansatzes such as unitary coupled cluster (UCC). The expectation value of an operator $\hat{P}$ can also be easily obtained by standard measurement techniques such as direct measurement on qubits corresponding to orbital indices, or alternatively the widely used Hadamard test as illustrated in Figure~\ref{fig::circ-cvqe}b.
    
    With the orbital-to-qubit mapping, calculating wave function based properties such as energy or reduced density matrices is straightforward. However, the number of qubits required to simulate the wave function itself has a linear dependence on the number of basis functions, which prohibits simulations of larger systems or the use of large basis sets on NISQ devices. For example, a simulation of H$_2$ using cc-pVTZ basis set requires 56 qubits which are already more than the qubits used in any quantum computing experiment for chemical systems\cite{Huggins_QMC_EXPERI_2022}. A moderate calculation for a typical periodic system using gaussian basis set requires approximately $10^3$ qubits\cite{McClain_EOMCC_2017, Liu_PBC_2020, FAN_PBC_2021}, and this number can be further increased by 10-100 times if the plane-wave basis set is used\cite{Liu_BASIS_REV_2022}, leading to a more significant shortage in qubit resources.

	\subsection{Matrix Product State}
    The requirement for qubit resources is generally recognized as \textit{space complexity} of a quantum algorithm. On a classical computer, space complexity is also an important issue. In traditional quantum chemistry, the electronic wave function is first approximated as a single Slater determinant which is easy to calculate. Then, electron correlation is recovered by including more determinants. The number of all possible determinants increases exponentially with the number of orbitals. However, as a consequence of locality, weights of these determinants are likely to be parameterized by only a small amount of information proportional to system size\cite{Chan_DMRG_CHEM_2011}. By encoding locality using a connected graph of tensors, tensor network states can be used to reduce computational overhead.\cite{Orus_MPS_PEPES_2014, Schollwock_DMRG_MPS_2011, Verstraete_TN_2008, Cirac_TP_2009, Shi_TTN_2006, Murg_TTN_2010, Verstraete_PEPS_2006} MPS is one typical class of tensor network states\cite{Clark_UNIFY_TN_2018} which have been widely used to solve quantum chemistry problems.

    The electronic wave function can be represented by a CI expansion:
	\begin{equation}
		\left| \Psi \right \rangle = \sum_{i_1 i_2 \ldots i_N} {c_{i_1 i_2 \ldots i_N} |i_1 i_2 \ldots i_N \rangle},
	\end{equation}
	where $N$ is the number of spin orbitals, and $|i_1 i_2 \ldots i_N\rangle$ is the computation basis marked by binary number string $i_1 i_2 \ldots i_N$. The coefficients $\{c_{i_1 i_2 \ldots i_N}\}$ form an $N$-dimensional tensor which contains $2^N$ amplitudes. The MPS ansatz factorizes this rank-$N$ coefficient tensor into lower rank tensors $\{^k T\}$ which can be written as:
	\begin{equation}
		\label{eq::mps-rep}
		c_{i_1 i_2 \ldots i_N} = \sum_{u_0\ldots u_N} {{^1T^{i_1}_{u_0 u_1}} {^2T^{i_2}_{u_1 u_2}} \ldots {^NT^{i_N}_{u_{N-1} u_{N}}} }
	\end{equation}
    where $\{^k T^{i_k}_{u_{k-1} u_{k}}\}$ represents elements of the rank-$3$ tensor ${^k T}$ at the $k$-th site, with $i_k$ called the \textit{physical} index and $u_k$ the \textit{auxiliary} index. The maximum size of the auxiliary indices is defined as the \textit{bond dimension} of the MPS, which is denoted as $D=\max_{0\le k \le N}{\{u_k\}}$.
    In classical calculations, a canonical form of the MPS is often helpful to make the algorithm numerically more stable. A left-canonical MPS satisfies
	\begin{equation}
		\sum_{i_k} {\sum_{u_{k-1}} {({^k T^{i_k}_{u_{k-1} u^{'}_{k}}})^{*} \times {^k T^{i_k}_{u_{k-1} u_{k}}}} } = \delta_{u^{'}_{k} u_{k}}
	\label{eq::mps-left}
	\end{equation}
	for every $0\le k \le N$. Similarly, the right-canonical form is ensured if
	\begin{equation}
		\sum_{i_k} {\sum_{ u_{k}} {{^k T^{i_k}_{u^{'}_{k-1} u_{k}}} \times ({^k T^{i_k}_{u_{k-1} u_{k}}})^{*} } } = \delta_{u^{'}_{k-1} u_{k-1}}.
	\label{eq::mps-right}
	\end{equation}

    Using the MPS formalism, the orbitals are encoded into tensors with size $\mathcal{O}(D^{2})$, and the $N$-orbital electronic wave function is parameterized using a total of $\mathcal{O}(D^{2} \times N)$ parameters which lowered the space complexity from exponential to polynomial. The bond dimension $D$ determines the expressibility of an MPS, and it significantly influences the performance of MPS-based algorithms such as DMRG. If $D$ is allowed to grow exponentially, the MPS can represent any wave function using Equation~\ref{eq::mps-rep}. A wave function is said to be efficiently represented by an MPS if $D$ grows sub-exponentially with the system size. Constant bond dimension can be used if the system satisfies the one-dimensional area law, which is however usually violated by electronic Hamiltonians. Using the MPS formalism combined with proper orbital optimization methods, the DMRG algorithm is capable of surpassing the gold-standard CCSD(T) for typical chemical systems\cite{Eriksen_BENZENE_2020, Hachmann_DMRG_H_2006, Wouters_DMRG_POLYENE_2014} and pushes the scale of high-level ab-initio calculations up to 200 orbitals. However, the high bond-dimension requirement still makes a DMRG simulation prohibitive for large systems.

    \begin{figure}[]
    		\subfloat[]{\includegraphics[width=0.62\linewidth]{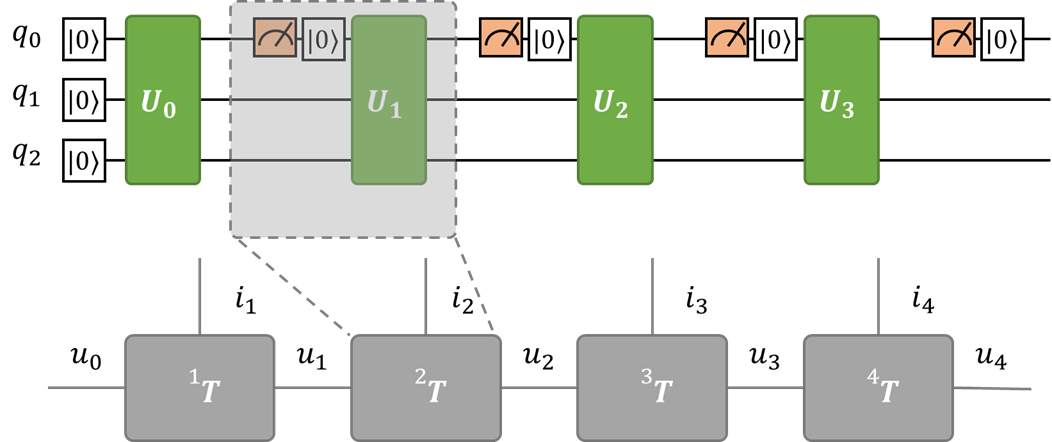}} \\
    		\subfloat[]{\includegraphics[width=0.62\linewidth]{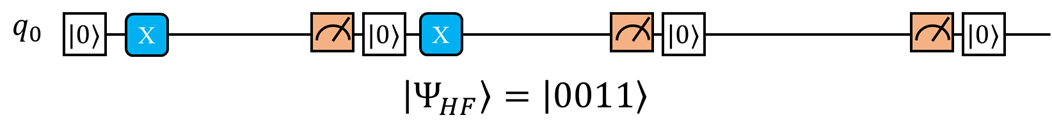}} \\
    		\subfloat[]{\includegraphics[width=0.62\linewidth]{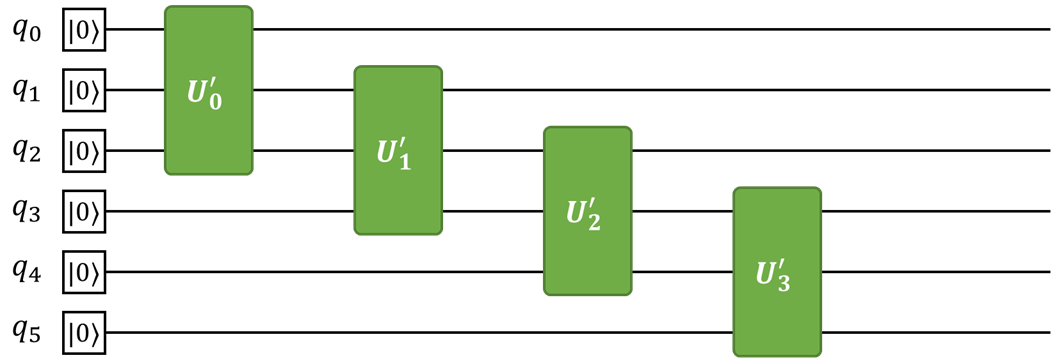}} \\
    		\caption{(a) A schematic QCMPS circuit.  The top qubit corresponds to the \textit{physical} indices $\{i_{k}\}$. A circuit block encoding a rank-3 MPS site tensor is marked with a dashed rectangle.  (b) A Hartree-Fock state represented by a QCMPS circuit. (c) The equivalent global circuit of (a), where \{\texttt{U}$_{i}$\} and \{\texttt{U}$_{i}^{'}$\} differs only by a few \texttt{SWAP} gates.}
    		\label{fig::circ-qmps}
    \end{figure}

    \subsection{Quantum Circuit MPS Ansatz}\label{sec::qcmps}
    The main idea of QCMPS is using quantum circuit blocks to represent the rank-3 tensors in MPS. Then, the number of blocks is determined by the number of orbitals or in other words the size of the system. The size of the block and thus the number of qubits is determined by the bond dimension instead of the number of basis functions. As shown in Figure~\ref{fig::circ-qmps}a, in a QCMPS with $N_q$ qubits, a circuit block corresponds to a specific orbital, which is similar to a rank-3 tensor in classical MPS with a bond dimension $2^{N_q-1}$. Within such a framework, the wave function represented by an MPS is encoded into a quantum circuit. Typically, the Hartree-Fock wave function can be represented with a $D=1$ QCMPS using only a single qubit (Figure~\ref{fig::circ-qmps}b).

    The orbital indices $\{i_{k}\}$ in the rank-3 tensors are not involved in the horizontal contraction of MPS. Accordingly, the $q_0$ qubit which corresponds to the orbital index should be reset after each circuit block. Such a qubit reset or sometimes called qubit reuse technique is realized by a mid-circuit measurement followed by conditioned \texttt{X} gate, as marked by a measurement operation plus a $|0\rangle$ in Figure~\ref{fig::circ-qmps}. It is also possible to avoid mid-circuit measurement by using more qubits to construct an equivalent \textit{global} version of QCMPS. As shown in Figure~\ref{fig::circ-qmps}c, in global QCMPS circuit, the entanglement blocks $\{U_{i}^{'}\}$ are consecutively applied on $\{q_0, q_1, q_2\}$, $\{q_1, q_2, q_3\}$, $\{q_2, q_3, q_4\}$, $\{q_3, q_4, q_5\}$ respectively, and the output states are on the first 4 qubits $\{q_0, q_1, q_2, q_3\}$ at the end of the circuit. The blocks in QCMPS and global QCMPS differs only by a few \texttt{SWAP} gates.

\begin{figure}[htbp]
        \subfloat[]{\includegraphics[width=0.62\linewidth]{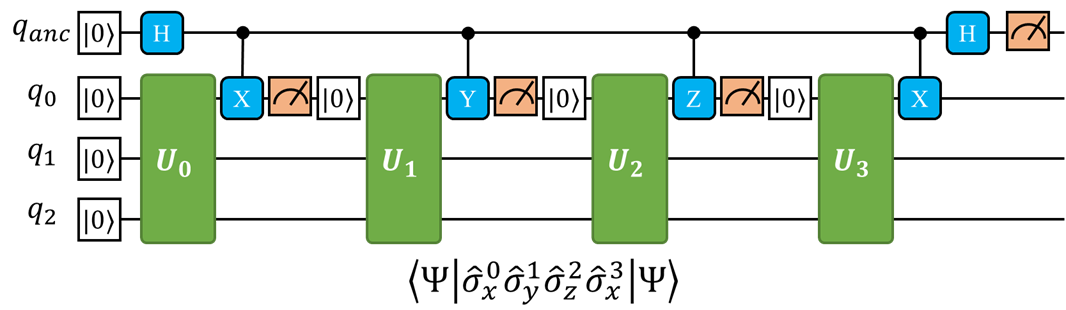}} \\
        \subfloat[]{\includegraphics[width=0.62\linewidth]{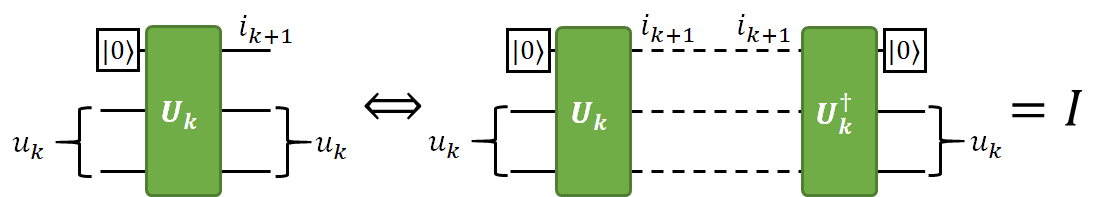}}
        \caption{ (a) Circuit to calculate the expectation value of a Pauli string with a QCMPS using Hadamard test. An ancillary qubit $q_{anc}$ is introduced. (b) QCMPS is a right-canonicalized MPS. }
        \label{fig::qcmps-cmps}
    \end{figure}

    Calculating expectation values of physical quantities on the top of QCMPS is slightly different from conventional orbital-to-qubit VQE ansatzes. Figure~\ref{fig::qcmps-cmps}a gives an example of measuring the expectation value of a 4-qubit operator using Hadamard test. The controlled gates are always applied on $q_{0}$ which is to be reset after each circuit block, and the measurement results from the ancillary qubit $q_{anc}$ are used to calculate expectation values.

    Since the evolution of quantum states can be regarded as tensor contractions, QCMPS corresponds to a right-canonical MPS as defined in Equation~\ref{eq::mps-right}, as illustrated in Figure~\ref{fig::qcmps-cmps}b. Due to the sparse parameterization, the variational expressibility of QCMPS is not only determined by the number of qubits (bond dimension) but also constrained by specific implementation of circuit blocks. For example, a circuit block with all qubit pairs entangled is expected to perform better than a nearest-neighbour-entangled block, while a general $N_{q}$-qubit unitary containing $\mathcal{O}(4^{N_{q}})$ parameters is expected to be over redundant. Therefore, choosing a suitable circuit structure is important in constructing QCMPS for chemical applications.

\begin{figure}[H]
		\subfloat[]{\includegraphics[width=0.60\linewidth]{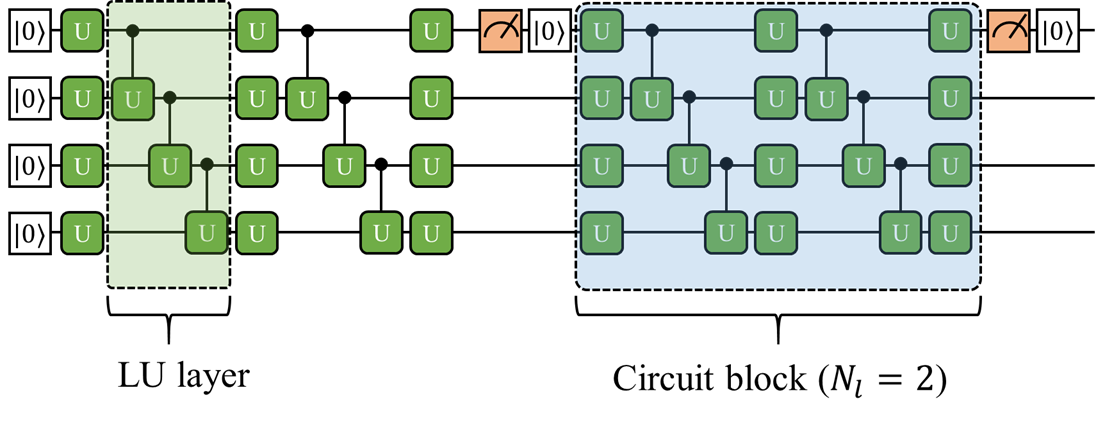}} \\
		\subfloat[]{\includegraphics[width=0.60\linewidth]{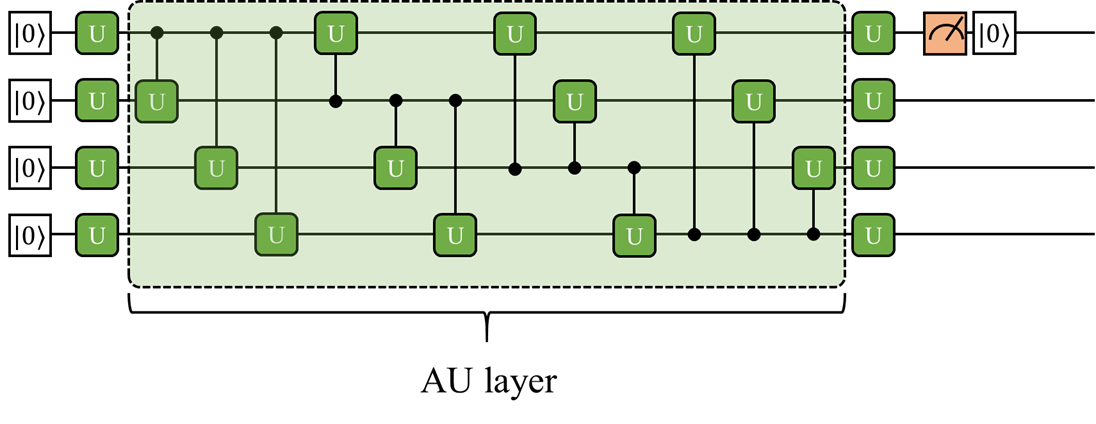}} \\
		\caption{QCMPS circuit constructed with (a) \texttt{LU} or (b) \texttt{AU} layers.}
		\label{fig::circ-block}
\end{figure}
	
	Generally, a circuit block contains some single-qubit gates and $N_{l}$ entangling layers (Figure~\ref{fig::circ-block}). In this work, two general types of entangling layers are studied, including linearly entangled layers which entangles neighbouring qubits consecutively (termed as \texttt{LU}) and fully entangled layers which entangles all qubit pairs (termed as \texttt{AU}). An extension of \texttt{AU} terms as \texttt{G2} is also used by replacing the 3-parameter controlled-\texttt{U} gate with a generic $4\times4$ unitary matrix, which can be decomposed using approximately 19 elementary one- or two-qubit quantum gates.\cite{Mikko_UNIV_GATE_2004} The notation $T\text{-}N_{q}$($N_{l}$) is used to represent an $N_{q}$-qubit QCMPS with $N_{l}$ layers of type $T$ structure in each block, where $T\in \{\texttt{LU}, \texttt{AU}, \texttt{G2}\}$.

    \subsection{Orbital Localization and Orbital Interaction}
    In classical MPS-based algorithms such as DMRG, localized orbitals are often used to lower the requirement on bond dimension\cite{Mitrushchenkov_DMRG_LO_2012}. It is thus expected that a proper orbital localization method can also be helpful for QCMPS. Starting from a set of canonical molecular orbitals (CMO) which are expressed as linear combinations of atomic orbitals:
	\begin{equation}
	    \phi_{s}(\bm{r}) = \sum_{\mu}^{N} {\chi_{\mu}(\bm{r}) C_{\mu s}},
	\end{equation}
	a set of localized orbitals can be expressed as linear combinations of the CMOs:
	\begin{equation}
		\tilde{\phi_{p}}(\bm{r}) = \sum_{s}^{N} {\phi_{s}(\bm{r}) U_{s p}},
	\label{eq::orb-rot-on-cmo}
	\end{equation}
	where the coefficients $\{U_{sp}\}$ form a unitary matrix. Several orbital localization approaches have been proposed, such as the natural atomic orbitals (NAO)\cite{Reed_NAO_1985}, Foster-Boys\cite{Foster_BOYS_1960}, Pipek-Mezey\cite{Pipek_PM_1989} and Edmiston-Ruedenberg \cite{Edmiston_ER_1963} methods. 
	Orbital transformation can also be regarded as applying an orbital rotation operator $\hat{R}$ on the wave function:
    \begin{equation}
        \begin{aligned}
            \hat{R} = \exp{(-\hat{\kappa})},
        \end{aligned}
    \end{equation}
    where $\hat{\kappa}$ can be written as:
    \begin{equation}
        \begin{aligned}
            \hat{\kappa} = \sum_{pq} {\kappa_{pq} \hat{a}^{\dagger}_{p} \hat{a}_{q}}.
        \end{aligned}
    \label{eq::kappa}
    \end{equation}
    The anti-hermitian matrix $\kappa$ is the generator of $U$ in Equation~\ref{eq::orb-rot-on-cmo}, which satisfies $U = \exp{(-\kappa)}$. This orbital rotation operator $\hat{R}$ is similar to a unitary coupled-cluster wave function with only single excitations and can be implemented as a quantum circuit ansatz.\cite{Sokolov_QOO_2020}
	
    In different orbital basis, there are generally distinct interactions between orbitals. Using quantum information theory, the orbital interaction can be characterized based on von Neumann entropies:\cite{Boguslawski_ORB_ENT_2015, Rissler_ORB_ENT_2006}
	\begin{equation}
	    I_{pq} = \frac{1}{2} (s^{2}_{pq} - s^{1}_{p} - s^{1}_{q})(1 - \delta_{pq}).
	\end{equation}
	where $\{p, q|p\neq q\}$ are indices of orbitals, and the one- or two-orbital entropies $s^{1}_{p}$ or $s^{2}_{pq}$ can be calculated from eigenvalues of one- or two-orbital reduced density matrices $\{\omega_{\alpha_{p}}\}$ and $\{\omega_{\alpha_{pq}}\}$ :
	\begin{equation}
	    \begin{aligned}
	        s^{1}_{p} = -\sum_{\alpha_{p}=1}^{4} {\omega_{\alpha_{p}} \ln{\omega_{\alpha_{p}}}}, \\
	        s^{2}_{pq} = -\sum_{\alpha_{pq}=1}^{16} {\omega_{\alpha_{pq}} \ln{\omega_{\alpha_{pq}}}}.
	    \end{aligned}
	\end{equation}
	A large value of $I_{pq}$ indicates that interactions between orbital $p$ and $q$ are significant. If $I_{pq}$ form a diagonal dominant matrix, the wave function is locally entangled and can thus be simulated using an MPS with a small bond dimension. In contrast, if there are large off-diagonal $I_{pq}$ elements, usually a larger or even exponentially growing bond dimension is required. It should be noted that geometrically more localized orbitals does not necessarily result in a more MPS-friendly wave function. As will be discussed in Section~\ref{seq::result}, orbital localization in certain cases tends to give an orbital interaction matrix with even more scattered pattern compared to the case of Hartree-Fock orbitals.

	\section{Numerical Results}\label{seq::result}

    We first study the performance of different circuit structures. Then, the influence of orbital localization is studied by calculating the orbital interaction matrices. Finally, we use linear hydrogen chains to study the expressibility of QCMPS by comparing with DMRG results. All calculations are performed using the Q$^2$Chemistry package\cite{FAN_Q2CHEM_2022} interfaced with several open-source software, including PySCF\cite{pyscf} for one- and two-electron integrals and SciPy\cite{scipy} for the Broyden-Fletcher-Goldfarb-Shanno (BFGS) optimizer.
	The STO-3G basis set is used for H$_4$, H$_6$ and H$_8$, and STO-6G is used for the hydrogen chains. Full configuration interaction (FCI) results are obtained by direct diagonalization of the qubit Hamiltonian. DMRG calculations are performed using an in-house developed code based on PyTorch.\cite{pytorch} The optimization of QCMPS is carried out with constrains on particle number and total spin.

\begin{figure}[]
		\subfloat[]{\includegraphics[width=0.3\linewidth]{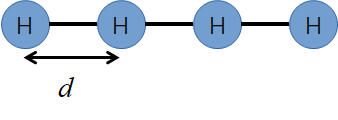}} \\
		\subfloat[]{\includegraphics[width=0.16\linewidth]{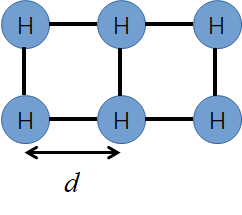}}
		{                }
		\subfloat[]{\includegraphics[width=0.14\linewidth]{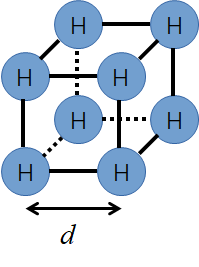}} \\
        \subfloat[]{\includegraphics[width=0.45\linewidth]{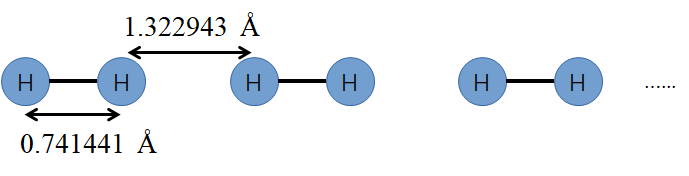}}
		\caption{Geometries of (a) linear H$_4$, (b) rectangular H$_6$, (c) cubic H$_8$ molecules and (d) one-dimensional hydrogen chain. H-H distance $d$ in H$_4$, H$_6$ and H$_8$ molecules is $2.0$~\r{A}.}
		\label{fig::h4_h6_h8_mol}
	\end{figure}

\begin{table}[]
	    \centering
	    \begin{tabular}{cccccccc}
\hline 
 $\Delta E$ & \texttt{LU}-4(1) & \texttt{LU}-4(3) & \texttt{LU}-4(7) & \texttt{AU}-4(1) & \texttt{AU}-4(3) & \texttt{G2}-4(1) & \texttt{G2}-4(3) 
 \\
\hline
H$_4$ & 201.893 & 1.302 & 1.302 & 1.301 & 1.301 & 1.301 & 1.301 
 \\
H$_6$ & 295.652 & 158.263 & 126.361 & 127.359 & 123.361 & 123.346 & 115.449 
 \\
H$_8$ & 363.896 & 269.661 & 232.374 & 244.085 & 226.038 & 228.248 & 211.035 
 \\
\hline 
$N_{p}$ & \texttt{LU}-4(1) & \texttt{LU}-4(3) & \texttt{LU}-4(7) & \texttt{AU}-4(1) & \texttt{AU}-4(3) & \texttt{G2}-4(1) & \texttt{G2}-4(3) 
\\
\hline 

H$_4$ & 264 & 600 & 1272 & 480 & 1248 & 1664 & 4992 
 \\
H$_6$ & 396 & 900 & 1908 & 720 & 1872 & 2496 & 7488 
 \\
H$_8$ & 528 & 1200 & 2544 & 960 & 2496 & 3328 & 9984 
\\

\hline \hline 
	         
	    \end{tabular}
	    \caption{Error with respect to FCI in kcal/mol ($\Delta E$) and number of variational parameters in QCMPS circuits ($N_{p}$) for H$_4$, H$_6$ and H$_8$ molecules. }
	    \label{tab::h4_h6_h8}
	\end{table}

    
    Effects of circuit structures are studied using linear H$_4$, rectangular H$_6$ and cubic H$_8$ molecules (Figure~\ref{fig::h4_h6_h8_mol}a-c). Since the purpose of this test is to study the performance of different types of circuit blocks, we use four qubits $N_{q}=4$ for all molecules, which does not converge the results to chemical accuracy. As shown in Table~\ref{tab::h4_h6_h8}, with a fixed number of qubits, the accuracy can be improved by increasing the number of layers $N_{l}$ before it goes to saturation. For the linear H$_4$ molecule, three \texttt{LU} layers are capable of approaching the accuracy of a single \texttt{AU} layer, which already makes \texttt{LU} structure less favorable regarding the number of parameters. In the case of non-linear molecules H$_6$ and H$_8$, seven \texttt{LU} layers are required for an accuracy comparable to a single-layer \texttt{AU} structure. In these cases, to achieve similar accuracy, \texttt{LU} structures leads to a two- to three-fold larger number of parameters to be optimized compared to \texttt{AU} structures. These results indicate that \texttt{AU} blocks are more efficient in capturing electron correlations in complex systems than \texttt{LU}. Therefore, we mainly use the fully-entangled \texttt{AU} structure in our QCMPS circuits. If \texttt{G2} blocks are used, the accuracy can be further improved on top of the \texttt{AU} circuit with the same $N_{l}$ but more parameters. 
	
	\begin{figure}[]
		\subfloat[]{\includegraphics[width=0.3\linewidth]{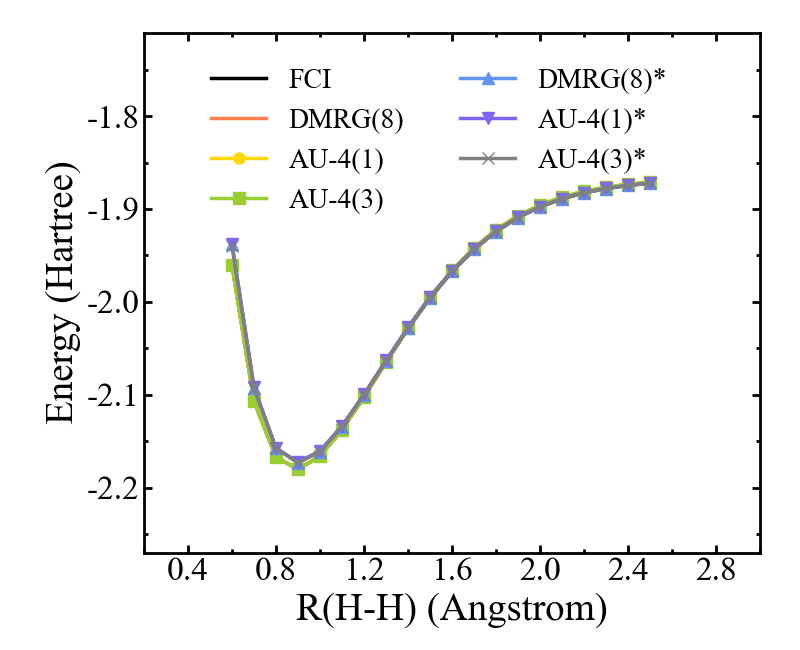}}
		\subfloat[]{\includegraphics[width=0.3\linewidth]{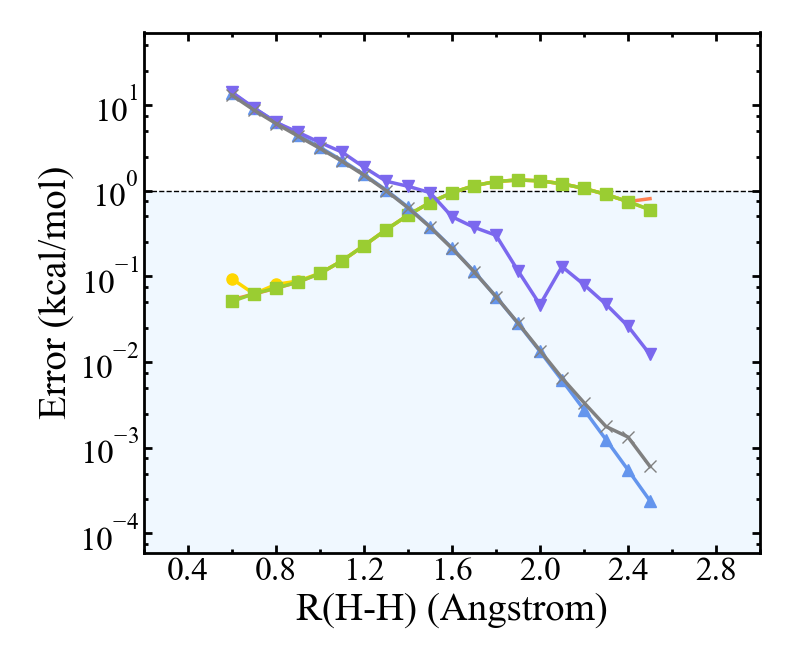}}
		\caption{(a) Potential energy curve of H$_4$ molecule with R(H-H) ranging from 0.6 to 2.5 \r{A} and (b) the corresponding errors with respect to FCI energies. Asterisk marks indicate that NAOs are used, otherwise CMOs are used. Shaded area represents the chemical accuracy ($\Delta E \le 1.0$ kcal/mol).}
		\label{fig::qmps-h4}
	\end{figure}

    Another flexibility in the QCMPS model is the basis set to construct the wave function. Typically, Hartree-Fock orbitals are commonly used to construct correlated wave function. However, such delocalized CMOs can be suboptimal for MPS-like ansatzes. It is therefore interesting to further study the influence of orbital localization. Four-qubit QCMPS results for the H$_4$ molecule obtained using CMOs and NAOs are shown in Figure~\ref{fig::qmps-h4}. DMRG results with an equivalent bond dimension $D=8$ are also given as a reference. For the strongly correlated region where bond length R(H-H) is large, using NAOs for QCMPS significantly outperforms canonical HF orbitals and is able to reduce the errors by orders of magnitude. However, NAO is not universally preferable across the potential energy surface, as shown for the results with R(H-H)$< 1.50$ \r{A}ngstrom.

    \begin{figure}[htbp]
        \centering
        \subfloat[]
        {
            \begin{minipage}[t]{0.45\linewidth}
                \centering
                \includegraphics[width=0.637\linewidth]{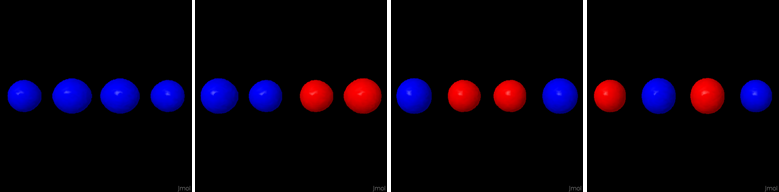} \\
                \includegraphics[width=0.84\linewidth]{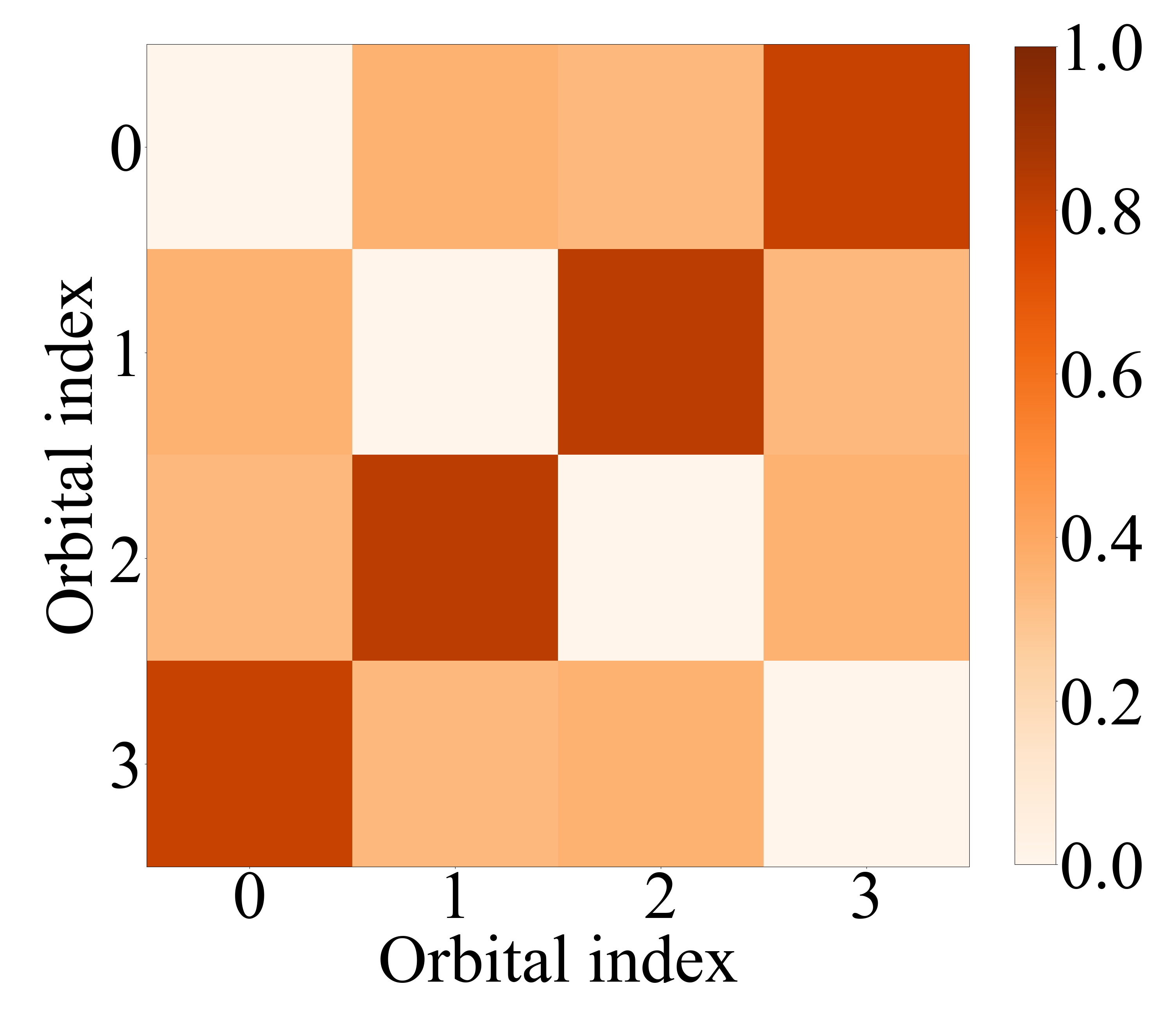}
            \end{minipage}
        }
        \subfloat[]
        {
            \begin{minipage}[t]{0.45\linewidth}
                \centering
                \includegraphics[width=0.637\linewidth]{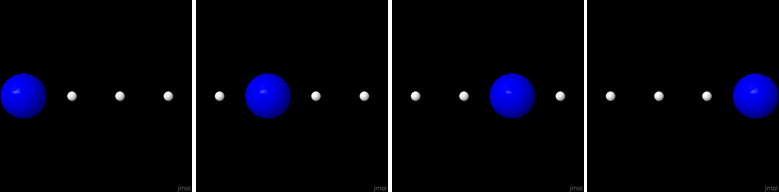} \\
                \includegraphics[width=0.84\linewidth]{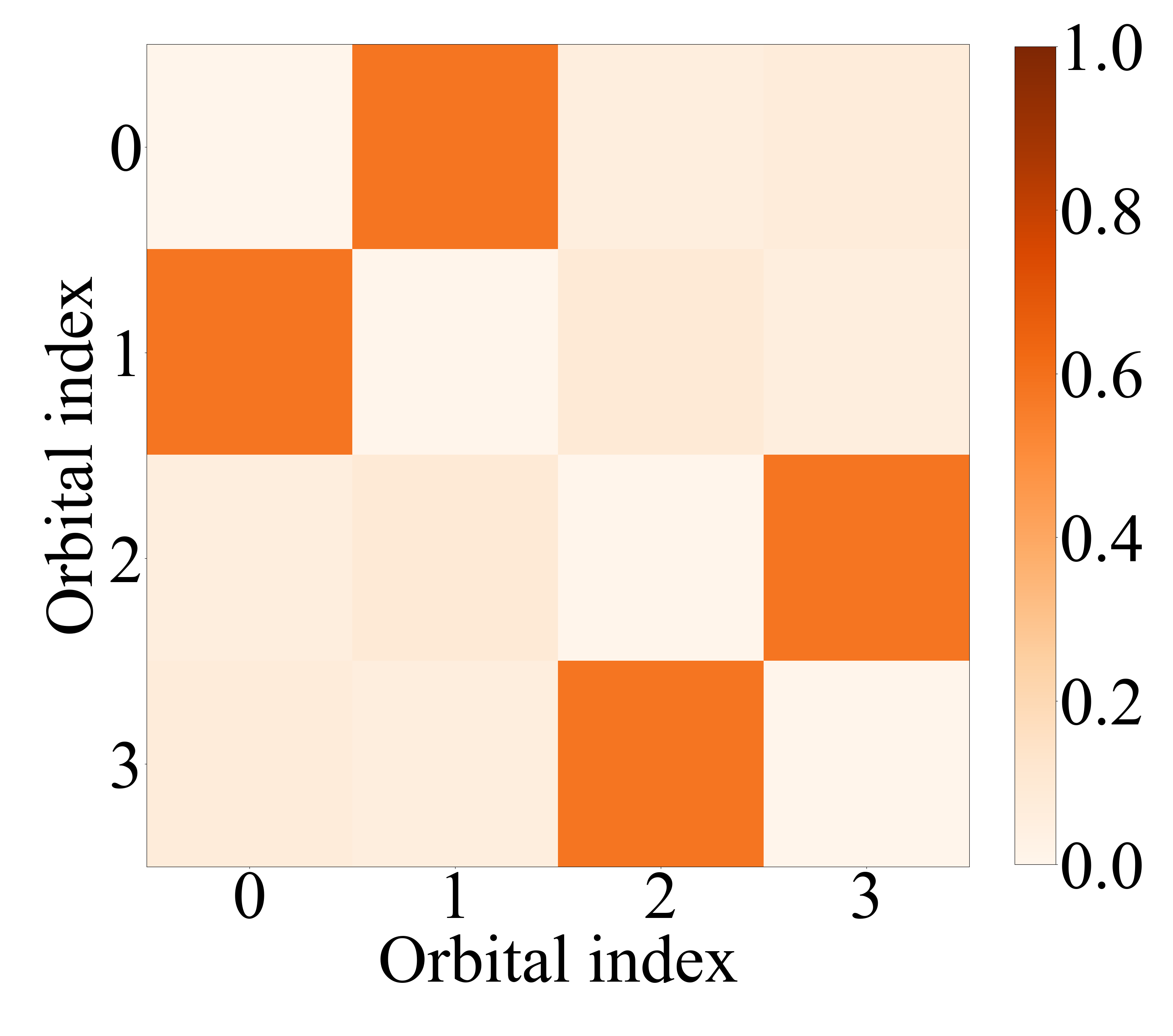}
        \end{minipage}
        }

        \subfloat[]
        {
            \begin{minipage}[t]{0.45\linewidth}
                \centering
                \includegraphics[width=0.637\linewidth]{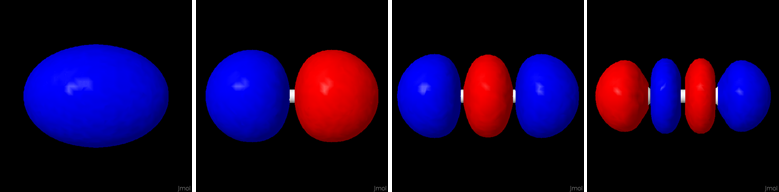} \\
                \includegraphics[width=0.84\linewidth]{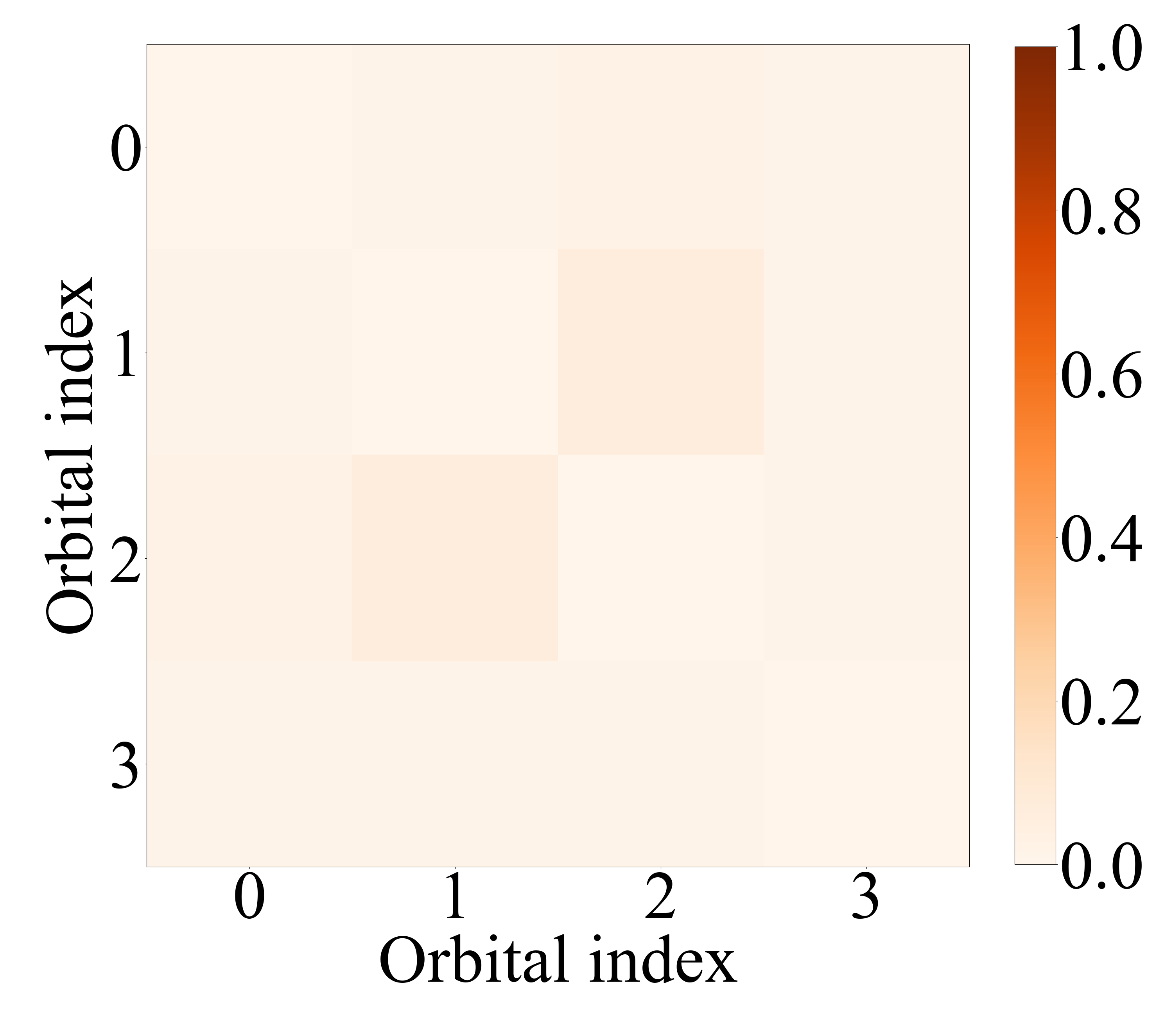}
            \end{minipage}
        }
        \subfloat[]
        {
            \begin{minipage}[t]{0.45\linewidth}
                \centering
                \includegraphics[width=0.637\linewidth]{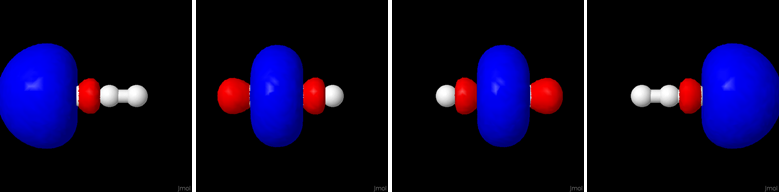} \\
                \includegraphics[width=0.84\linewidth]{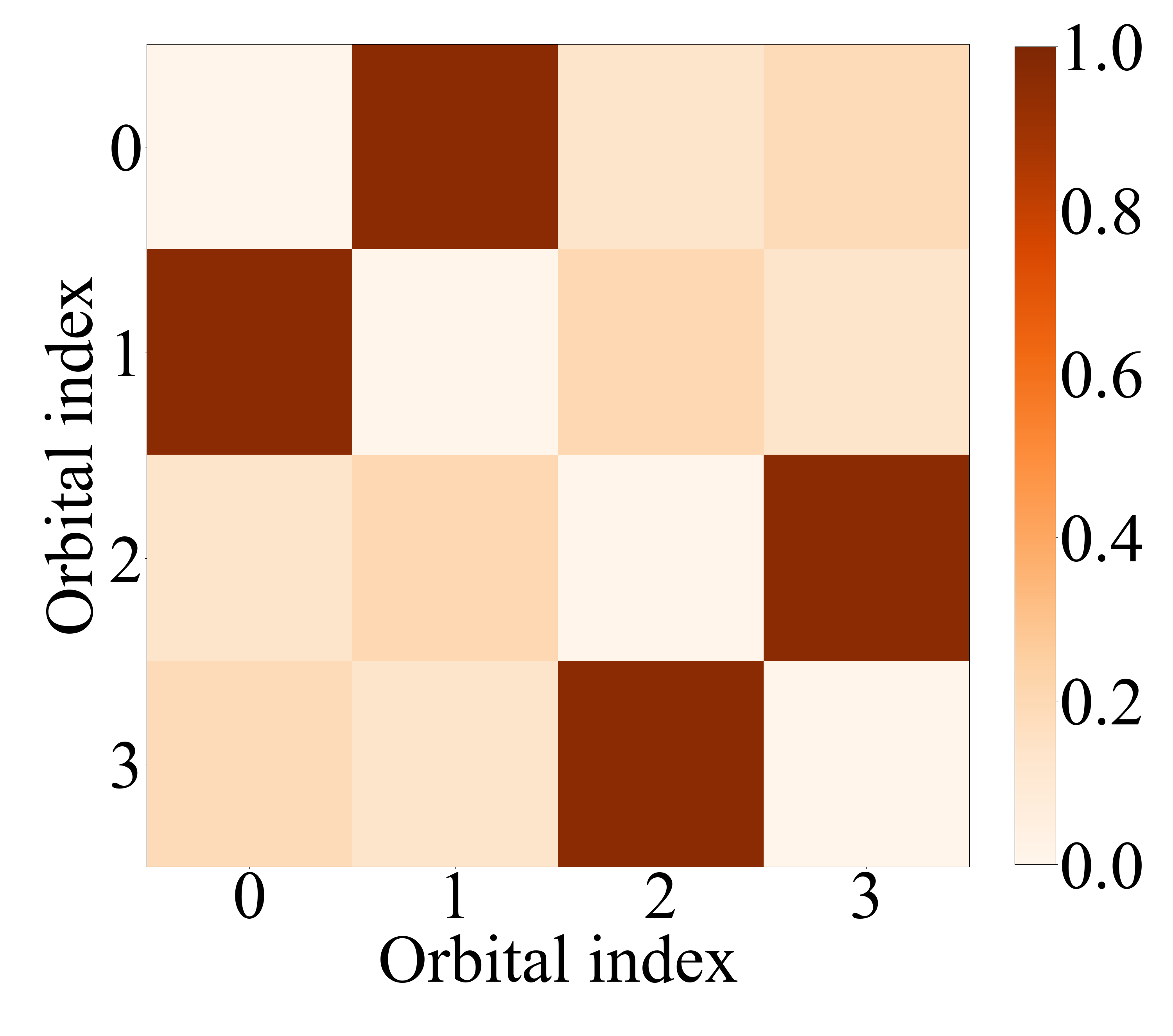}
        \end{minipage}
        }

        \caption{Orbitals and the corresponding orbital interaction matrices for QCMPS wave functions of the H$_4$ molecule. (a) CMOs and (b) NAOs at R(H-H)=2.5 \r{A}. (c) CMOs and (d) NAOs at R(H-H)=0.6 \r{A}.}
        \label{fig::orb_ent_h4}
    \end{figure}

    To understand such a degradation at small bond lengths, we calculate orbital interaction matrices for ground state wave functions of the H$_4$ molecule at R(H-H)$=$0.6 and 2.5 \r{A}. As shown in Figure~\ref{fig::orb_ent_h4}, NAOs successfully concentrate entanglement onto neighbouring orbitals at R(H-H)=2.5\r{A}, which leads to a QCMPS-friendly wave function. However, at R(H-H)=0.6\r{A}, the interactions between NAOs are still scattered and the values of $I_{pq}$ are even larger than CMOs even if the NAOs are geometrically more localized. Although NAOs greatly improve the accuracy of QCMPS at regions where static correlations are significant, they fail to give consistently accurate results across the potential energy curve. An inappropriate localization method can lead to large orbital entanglement and even scatters strongly interacting orbitals far apart, which makes the wave function beyond the expressibility of QCMPS with a small number of qubits. An optimal choice of orbital localization methods for such an MPS-inspired ansatz is usually system-specific and requires a benchmark for complex systems\cite{Yingjin_DMRG_LO_2013}. Previous studies show that the general structure of $I_{pq}$ are barely affected by the bond dimension\cite{Rissler_ORB_ENT_2006}. Therefore, in QCMPS simulations, it is possible to optimize orbital rotation parameters ($\{\kappa_{pq}\}$ in Equation~\ref{eq::kappa}) using a small $N_{q}$ before an accurate calculation.

\begin{table}[]
	    \centering
	    \begin{tabular}{ccccc}
        \hline 
        Number of orbitals & 24 & 32 & 40 & 50 \\ 
        \hline 
        $N_{q}=2$ & 1  & 1  & 1  & 1    \\ 
        $N_{q}=3$ & 2  & 2  & 2  & 2    \\ 
        $N_{q}=4$ & 3  & 3  & 3  & 3    \\ 
        $N_{q}=5$ & 4  & 5  & 5  & 4    \\ 
        $N_{q}=6$ & 6  & 6  & 6  & 6    \\ 
        \hline \hline 
        	         
        \end{tabular}
        \caption{Number of layers $N_{l}$ used in the QCMPS simulations of hydrogen chains. The $N_{l}$ is selected such that the energy difference of \texttt{G2}-$N_{q}(N_{l})$ and \texttt{G2}-$N_{q}(N_{l}-1)$ is smaller than $2.0\times 10^{-5}$ Hartree.}
        \label{tab::g2}
\end{table}

\begin{figure}[]
    \includegraphics[width=0.45\linewidth]{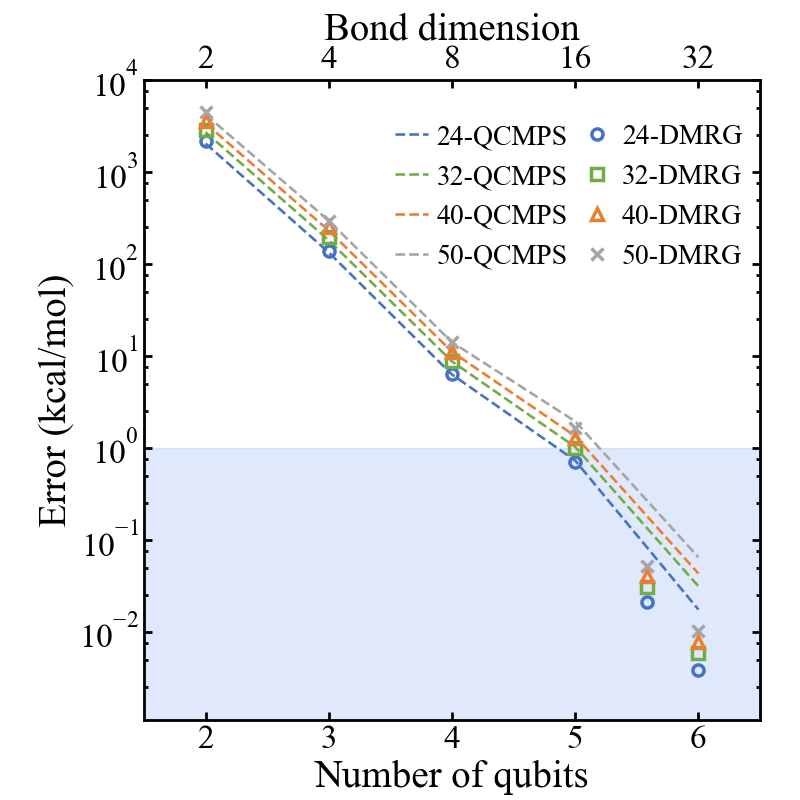}
    \caption{Error of QCMPS and DMRG simulations of H chains with respect to well converged DMRG energies obtained at a bond dimension of $D=128$. The numbers before 'QCMPS' and 'DMRG' give the number of orbitals in hydrogen chains. Shaded area represents the chemical accuracy ($\Delta E\le$1.0 kcal/mol).}
    \label{fig::h2-mol-chain}
\end{figure}

    To demonstrate the power of QCMPS, we first calculate the potential energy curve of H$_2$ with CMOs. The chemical accuracy can be readily obtained with an \texttt{AU}-2(1) circuit (Figure~S1) as expected. Then, the ground state energies of more complicated one-dimensional hydrogen chains with alternate H-H distances (Figure~\ref{fig::h4_h6_h8_mol}d) are calculated. Here, the number of orbitals $N_{o}$ equals to the number of hydrogen atoms. We consider four chains with $N_{o}=$ 24, 32, 40 and 50, respectively. Due to the excellent performance in the strongly correlated region, NAOs are used in these systems. As listed in Table~\ref{tab::g2}, adequate number of \texttt{G2} entanglement layers are used in the QCMPS circuit blocks. Therefore, the number of qubits $N_{q}$ becomes the only hyperparameter of the QCMPS model, which corresponds to a bond dimension $D = 2^{N_{q} - 1}$ in DMRG simulations. 

    As shown in Figure~\ref{fig::h2-mol-chain}, errors in QCMPS simulations systematically decrease with the number of qubits in a similar way as the errors decrease with the corresponding bond dimensions in the DMRG calculations. The result that QCMPS reaches a comparable accuracy to DMRG with bond dimension close to $2^{N_{q} - 1}$ is not trivial, since the rank-3 tensors are sparse with constrains from circuit structure in QCMPS, while they contains a full flexibility in DMRG. In the hydrogen chain case, such a difference between QCMPS and DMRG starts to have an effect only when the number of qubits and the corresponding bond dimension is already large and the chemical accuracy has been reached by both methods.

    In this example, we demonstrate that a large molecule can be simulated with QCMPS using only a very small number of qubits. For a quantum circuit, another important parameter with practical significance is the circuit depth. As shown in Table~\ref{tab::g2}, the required number of entanglement layers in this example is roughly equal to $N_{q}$, which leads to a quantum circuit of $\mathcal{O}(N_{q}^{3} \times N)$ gates where $N$ is the number of orbitals. In this way, QCMPS is not only qubit efficient but also gate efficient compared with commonly used ansatzes such as UCCSD. For example, in the 50-orbital case, the total number of elementary one- and two-qubit gates used in \texttt{G2}-6(6) is approximately $3.4\times 10^5$, which is orders of magnitude smaller than that of the first-order Trotterized UCCSD ($5.9\times 10^{8}$) which has a $\mathcal{O}(N^4)$ scaling in circuit depth.

	\section{Conclusion}
    In this study, we have presented a qubit-efficient QCMPS ansatz for simulating chemical systems. QCMPS encodes right-canonical MPS tensors into quantum circuit blocks. The number of qubits controls the bond dimension of the tensors and the number of circuit blocks is determined by the number of orbitals in the system. A fully-entangled circuit structure is preferred in QCMPS for molecular systems. QCMPS is efficient in qubit resource for large-scale simulations. As a demonstration, QCMPS can use a maximum number of 6 qubits to simulate linear hydrogen chains with up to 50 orbitals, reaching similar accuracies compared to those achieved by DMRG with an exponentially large bond dimension. The QCMPS ansatz presented in this study thus represents an attractive way to perform quantum simulations for large molecular systems with limited qubit resources.

    \begin{acknowledgement}
		This work is supported by the Innovation Program for Quantum Science and Technology (2021ZD0303306), the National Natural Science Foundation of China (21825302), the Fundamental Research Funds for the Central Universities (WK2060000018), and the USTC Supercomputing Center.
	\end{acknowledgement}

	\bibliography{citations.bib}

\end{document}


\maketitle

    \subsection{S1. Potential Energy Curve of H$_2$}
\begin{figure}[]
		\subfloat[]{\includegraphics[width=0.45\linewidth]{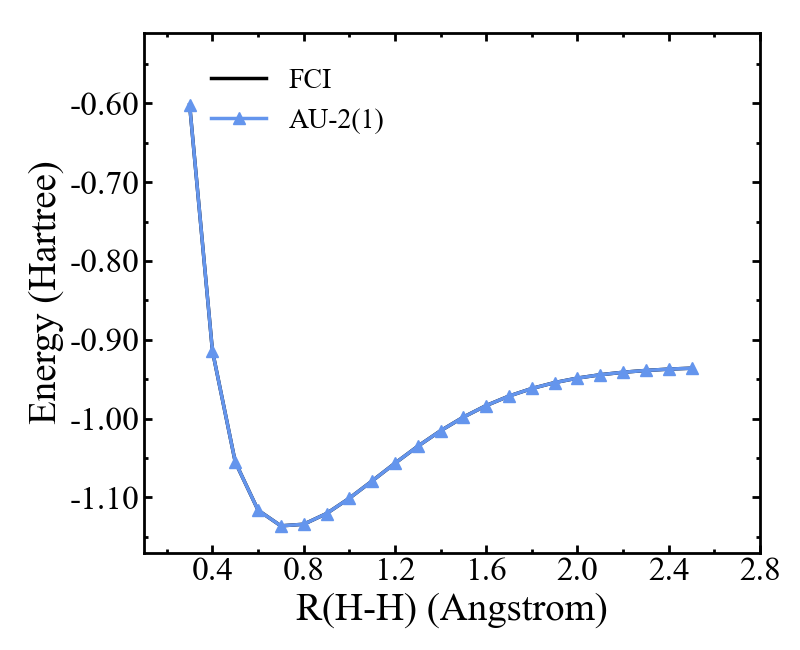}}
		\subfloat[]{\includegraphics[width=0.45\linewidth]{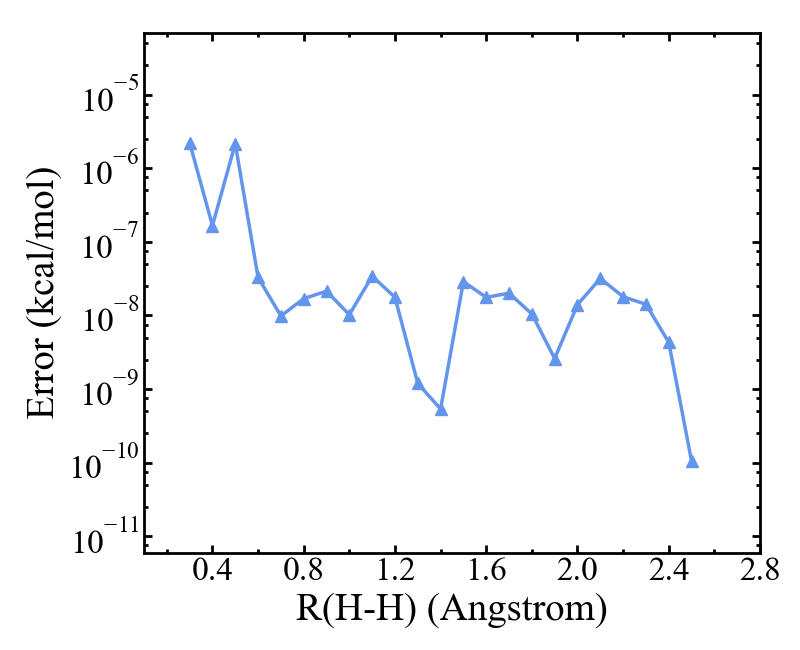}}
		\caption{The simulated (a) potential energy curve and (b) error with respect to FCI energy of H$_2$, using QCMPS ansatz with \texttt{AU}-2(1) blocks.}
		\label{fig::qmps-h2}
	\end{figure}

    The potential energy curve of H$_2$ calculated with bond length R(H-H) from 0.6 to 2.5 \r{A}ngstrom is shown in Figure~S\ref{fig::qmps-h2}. Canonical HF orbitals are used to perform the simulation. QCMPS can converge to chemical accuracy using \texttt{AU} block with $N_{q}=2$ and $N_{l}=1$. From the bond dimension point of view, the QCMPS ansatz used here is equivalent to a classical MPS with $D = 2$. 

